\documentclass[%
reprint,
superscriptaddress,
mathtools,amssymb,
aps,
floatfix,
]{revtex4-2}
\usepackage[ruled,vlined]{algorithm2e}
\usepackage{mathtools}
\usepackage[english]{babel}
\usepackage[caption=false]{subfig}
\usepackage{physics}
\usepackage[linguistics]{forest}
\usepackage{hyperref}
\usepackage{graphicx}%
\usepackage{dcolumn}%
\usepackage{bm}%
\usepackage{float}
\usepackage[pages=some,scale=1,angle=0,opacity=1]{background}
\usepackage{cleveref}
\crefname{equation}{Eq.}{Eqs.}
\Crefname{equation}{Equation}{Equations}
\crefname{figure}{Fig.}{Figs.}
\Crefname{figure}{Figure}{Figures}
\crefname{section}{Sec.}{Secs.}
\Crefname{section}{Section}{Sections}
\crefname{appendix}{Appendix}{Apps.}
\Crefname{appendix}{Appendix}{Apps.}
\crefname{paragraph}{Sec.}{Secs.}
\crefname{table}{Table}{Tables}

\colorlet{Ross}{red}
\colorlet{RS}{red}
\colorlet{Elie}{orange}
\colorlet{Jonathan}{purple}
\colorlet{Agustin}{blue}
\colorlet{AB}{teal}
\colorlet{Edit}{brown}

\begin{document}

	\title{Fast and differentiable simulation of driven quantum systems}

    \author{Ross Shillito}
    \email{Ross.Shillito@USherbrooke.ca}
    \address{Institut quantique \& D\'epartement de Physique, Universit\'e de Sherbrooke, Sherbrooke J1K2R1, Quebec, Canada}
    \author{Jonathan A. Gross}
    \thanks{Jarthurgross@google.com}
    \address{Institut quantique \& D\'epartement de Physique, Universit\'e de Sherbrooke, Sherbrooke J1K2R1, Quebec, Canada}
	\author{Agustin Di Paolo}
	\thanks{adipaolo@mit.edu}
    \address{Institut quantique \& D\'epartement de Physique, Universit\'e de Sherbrooke, Sherbrooke J1K2R1, Quebec, Canada}
	\author{\'Elie Genois}
    \address{Institut quantique \& D\'epartement de Physique, Universit\'e de Sherbrooke, Sherbrooke J1K2R1, Quebec, Canada}
	\author{Alexandre Blais}
	\address{Institut quantique \& D\'epartement de Physique, Universit\'e de Sherbrooke, Sherbrooke J1K2R1, Quebec, Canada}
    \address{Canadian Institute for Advanced Research, Toronto, M5G1M1 Ontario, Canada}
	
\begin{abstract} %
	The controls enacting logical operations on quantum systems are described by time-dependent Hamiltonians that often include rapid oscillations. In order to accurately capture the resulting time dynamics in numerical simulations, a very small integration time step is required, which can severely impact the simulation run-time. Here, we introduce a semi-analytic method based on the Dyson expansion that allows us %
	to time-evolve driven quantum systems much faster than standard numerical integrators. This solver, which we name \texttt{Dysolve}, efficiently captures the effect of the highly oscillatory terms in the system Hamiltonian, significantly reducing the simulation's run time as well as its sensitivity to the time-step size. Furthermore, this solver provides the exact derivative of the time-evolution operator with respect to the drive amplitudes. This key feature allows for optimal control in the limit of strong drives and goes beyond common %
	pulse-optimization approaches that rely on rotating-wave approximations. As an illustration of our method, we show results of the optimization of %
	a two-qubit gate using transmon qubits in the circuit QED architecture.
\end{abstract}
	
\date{\today}
	
\maketitle

\section{Introduction}
High-fidelity logical gates are paramount to the realization of useful quantum computation. It is important in %
the development of these gates that they be %
simulated with great precision to ensure that they meet the particularly strict requirements for fault-tolerant quantum computation. %
 Several techniques are used for simulating the time dynamics of quantum devices, %
including dynamical solvers such as Runge-Kutta integrators and direct matrix exponentiation~\cite{RungeKuttaMethods,19DubiousWays}. %
However, capturing the full time dynamics with the necessary accuracy requires integration methods with a sufficiently small time step. As a result, simulations can be computationally very expensive, even for relatively simple cases such as optimizing a two-qubit %
gate.

To ensure that simulations of quantum systems are feasible, approximations must be made. For example, %
a common approximation is to neglect counter-rotating terms within the rotating wave approximation (RWA) to greatly reduces the complexity of the simulation whilst capturing the dominant dynamics. This approximation renders the Hamiltonian time independent, which can then simply be exponentiated to obtain the propagator. However, effects such as the Bloch-Siegert shift which do not appear under a RWA need to be taken into account to accurately model the system~\cite{BlochSiegert1940}. 
In addition, %
many gate optimization methods, such as GRAPE, require the calculation of gradients, something which greatly adds to the complexity of the numerical calculations~\cite{KHANEJA2005296}. More specifically, when including the effects of the counter-rotating terms, %
there exists no simple derivative of %
time-ordered %
unitaries with respect to the drive amplitudes, and one must resort to approximating the %
gradients. Consequently, this approach %
may not converge to the optimal solution, which is problematic when targeting very high-fidelity gates.

In this work, we develop an algorithm %
based on a Dyson series expansion of the time ordered problem that %
addresses all of the above difficulties. In this approach, which we call \texttt{Dysolve}, the time-ordered unitary evolution operator is written as a product of time-independent operators which are weighted by the drive amplitudes and dynamical phases. This algorithm captures the full fast-oscillatory dynamics irrespective of the integration step size, %
thereby decreasing the complexity of the numerical problem. This also greatly decreases the simulation time in comparison to traditional integration-based solvers without loss of numerical precision. Importantly, this approach trivializes the derivative with respect to the drive strength, which can be calculated to an accuracy equivalent to the order of the Dyson series. Moreover, this approach is compatible with non-Hermitian dynamics, allowing for the simulation of open quantum systems.

We begin by introducing our approach in \cref{sec:GenericSinusoidal} in the case of a single, sinusoidal drive with a constant amplitude, and then extend the formalism to the case of multiple drives, accounting for filtering effects on envelope functions. We then define the \texttt{Dysolve} algorithm in \cref{sec:DysolveAlg}, and demonstrate its application to driven quantum systems with random %
drive envelopes. We proceed to apply our algorithm to the GRAPE optimization routine in \cref{sec:GRAPE}, and show as an example optimized two-qubit gates in the circuit QED architecture.%

\section{Oscillatory Drive Problem} \label{sec:GenericSinusoidal}

\subsection{Simple time-dependent Hamiltonians}

We begin by considering a simple time-dependent Hamiltonian with a cosinusoidally oscillating control drive term
\begin{equation}\label{eq:H(t)}
\begin{aligned}
\hat{H}(t) &= \hat{H}_0 + \hat{V}(t),
\quad  \hat{V}(t) = \hat{X} \cos(\omega t).
\end{aligned}
\end{equation}
Here,~$\hat{H}_0 = \sum_k \lambda_k \ketbra{k}{k}$ is a generic system Hamiltonian expressed in its eigenbasis, while $\hat{X}$ is a dipole operator that connects the eigenstates of $\hat{H}_0$ and which we take to account for the amplitude of the drive. As will become important later, we note that we are not using the RWA.

The propagator under $\hat{H}(t)$ for some time increment $\delta t$ takes the usual form of a time-ordered integral
\begin{equation}\label{eqn:TimeOrderedIntegral}
	\hat{U}(t,t+\delta t) = \mathcal{T}\exp(-i\int_{t}^{t+\delta t}dt'\hat{H}(t')),
\end{equation}
with $\mathcal{T}$ the time-ordering operator. Due to the fast oscillatory $\cos (\omega t)$ term, evaluating %
the propagator~$\hat{U}(t,t+\delta t)$ is a numerically challenging problem, and there exists few analytic solutions to even the simplest case of a driven two level qubit \cite{PhysRevLett.109.060401}. In special cases, %
one can invoke the RWA to remove the explicit time dependence from the Hamiltonian, thereby allowing for calculation of the propagator from matrix exponentiation directly \cite{PhysRevLett.98.013601}.

However, when the RWA cannot be used %
due to a breakdown of %
the approximation or because a greater degree of accuracy is needed, we propose using a truncated Dyson series. Consider \cref{eqn:TimeOrderedIntegral}, written using the definition of the time-ordering operator %
\begin{equation}
\begin{aligned}
&\hat{U}^{}(t,t+\delta t) =\\
&\sum_{n=0}^\infty (-i)^n \int_t^{t+\delta t} \int_t^{t_n}\cdots \int_t^{t_{2}}\hat{H}(t_n)\cdots \hat{H}(t_{1}) dt_1\cdots dt_n.
\end{aligned}
\end{equation}
It is useful to express this expansion in terms of powers of the drive operator $\hat{V}(t)$, forming the Dyson series
\begin{figure}
\includegraphics[width=1\columnwidth]{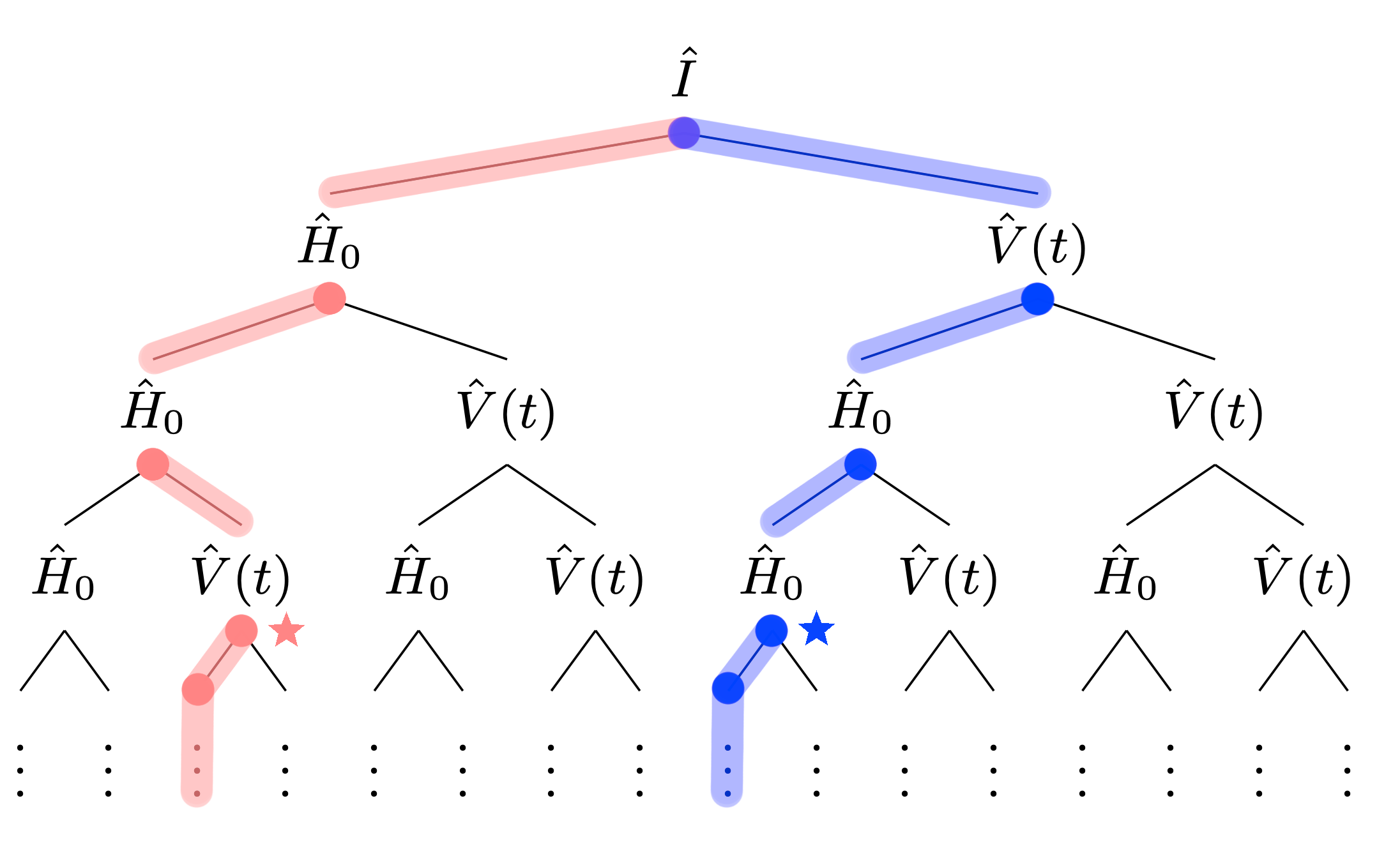}
\caption{Tree diagram showing the different branches of the time ordered integral, with two example branches highlighted.}
\label{fig:TREE}
\end{figure}
\begin{equation}
\label{eqn:OEdefn3}
\begin{aligned}
\hat{U}(t, t + \delta t) &= \sum_{n=0}^\infty \hat{U}^{(n)}(t, t + \delta t),\\
\end{aligned}
\end{equation}
where we have defined
\begin{equation}
\label{eqn:OEdefn4}
\begin{aligned}
\hat{U}^{(n)}(t, t + \delta t) &= \sum_{\bm{\omega}_n}\exp(i\sum_{p=1}^n\bm{\omega}_n[p] t) \hat{S}^{(n)}(\bm{\omega}_n,\delta t).
\end{aligned}
\end{equation}
Here,~$\bm{\omega}_n$ is an~$n$-vector whose entries are $\pm\omega$ %
originating from the decomposition of the $\cos(\omega t)$ of the control into complex exponentials, and $\bm{\omega}_n[p]$ is the $p$-th element of $\bm{\omega}_n.$ The sum over $\bm{\omega}_n$ implies a summation over all $2^n$ possible $\bm{\omega}_n$ vectors.

\Cref{eqn:OEdefn4} also introduces the Dyson series operator~$\hat{S}^{(n)}(\bm{\omega}_n,\delta t)$ which takes the form
\begin{equation} \label{eqn:FormalNthDysonOperatorsinglecomb}
    \hat{S}^{(n)}(\bm{\omega}_n,\delta t)  =\frac{1}{2^n}\sum_{\mathclap{\bm{m}\in\mathbf{Z}_+^{n+1}}} \hat{S}^{(n)}_{\bm{m}}(\bm{\omega}_n,\delta t),
\end{equation}
and which corresponds to a summation over all Dyson path operators of $n$-th order, where $\bm{m} = [m_n,...,m_0]$ with each index $m_i$ ranging from $0$ to infinity. These $n$-th order path operators are given by 
\begin{equation}\label{eqn:PathOperatorNthOrder}
\begin{aligned}
&\hat{S}^{(n)}_{\bm{m}}(\bm{\omega}_n,\delta t) =\\
&\int_0^{\delta t}\!\!\int_0^{t'_{M}}\!\!\!\!\!\cdots\!\int_0^{t'_2} (-i\hat{H}_0)^{m_n}\hat{X} (-i\hat{H}_0)^{m_{n-1}}\!\cdots\!\hat{X}(-i\hat{H}_0)^{m_0} \\
&\times (-i)^n  \prod_{p=1}^{n}\exp(i \bm{\omega}_n\left[p\right]t_{\iota(p)}) 
dt'_1\cdots dt'_M
\end{aligned}
\end{equation}
where we have introduced
\begin{equation}
	M = \sum_{i=0}^n m_i + n, \quad \iota(p) =  \sum_{j=0}^{p-1} m_{j}+p.
\end{equation}
Each %
$\hat{S}^{(n)}_{\bm{m}}(\bm{\omega}_n,\delta t)$ correspond to different ways to have $n$ contributions from the control $\hat{X}$ (i.e.~different terminated branches of the tree diagram), with the $m_i$'s labeling the number of applications of $\hat{H}_0$ before the subsequent application of $\hat{X}$. The total number of operators in the path operator is given by~$M$. To simplify the %
notation, we introduce an indexing function $\iota(p)$,which can be interpreted as the total number of $\hat{H}_0$ and $\hat{X}$ operators before the $p$-th application of the control $\hat{X}$.  These definitions can be visualized as in \cref{fig:TREE} where the red and blue branches  correspond to sets of path operators $\hat{S}^{(n)}_{[m_1,2]}(\bm{\omega}_n,\delta t)$ and $\hat{S}^{(n)}_{[m_1,0]}(\bm{\omega}_n,\delta t)$ respectively, with $m_1$  determined by where the path terminates.
For example, $\hat{S}^{(n)}_{[0,2]}(\bm{\omega}_n,\delta t)$ and $\hat{S}^{(n)}_{[2,0]}(\bm{\omega}_n,\delta t)$ terminate at the points in the branches marked by a star. 

Below, we give explicit expressions for these operators to zeroth and first order in the control. Building on these results, we then construct expressions that are easily amenable to efficient numerical evaluation to arbitrary orders.

\subsection{Evaluation to zeroth and first order}\label{sec:01stordereval}

The zeroth order corresponds to the leftmost branch of the tree diagram of \cref{fig:TREE} for which it is straightforward to obtain an explicit expression. Indeed, the path operator simply takes the form
\begin{equation}
\begin{aligned}
\hat{S}^{(0)}_{[m_0]}(\bm{0},\delta t) &= \int_{0}^{\delta t}\int_{0}^{t_{m_0}}\!\!\!\!\cdots \int_{0}^{t_{2}} (-i\hat{H}_0)^{m_0} dt_1\cdots dt_{m_0-1}dt_{m_0}\\
 &= \frac{(-i\hat{H}_0 \delta t)^{m_0}}{m_0!}.
\end{aligned}
\end{equation}
Summing all of the elements in the branch, we obtain
\begin{equation}
\begin{aligned}
\hat{S}^{(0)}(\bm{0},\delta t) = \sum_{m_0=0}^\infty \hat{S}^{(0)}_{[m_0]}(\bm{0},\delta t)
= e^{-i\hat{H}_0 \delta t},
\end{aligned}
\end{equation}
which corresponds, as expected, to the drift evolution of the Hamiltonian in the absence of drive. In a similar fashion, the path operator to first order takes the form
\begin{equation}
\label{eqn:PathOperators1stORder}
\begin{aligned}
&\hat{S}^{(1)}_{[m_1,m_0]}\left(\bm{\omega}_1,\delta t \right) = \int_0^{\delta t}\int_0^{t_{M}}\!\!\!\!\cdots\int_0^{t_2} (-i\hat{H}_0)^{m_1}\hat{X} (-i\hat{H}_0)^{m_{0}} \\
&\times \exp(\pm i[\omega] t_{\iota(1)}) dt_1\cdots dt_M,
\end{aligned}
\end{equation}
and therefore leads to the following %
Dyson series operator
\begin{equation} \label{eqn:Formal1stDysonOperator}
	\hat{S}^{(1)}(\bm{\omega}_1,\delta t) = -\frac{i\delta t}{2} \sum_{k,k'} f(\lambda_{k}\delta t,(\lambda_{k'}\mp\omega)\delta t) \langle {k} | \hat{X} | {k'} \rangle   |{k} \rangle \langle {k'}|.
\end{equation}
This operator weights the matrix elements of $\hat{X}$ by a function $f$ whose inputs are weighted eigenvalues $\{\lambda_k\}$ %
of the free Hamiltonian~$ \hat{H}_0$ with corresponding eigenstates~$\{\ket k\}$, %
where the second eigenvalue is shifted by the drive frequency $\bm{\omega}_1[1] = \pm \omega$. This function is defined as
\begin{equation} \label{eqn:1storderweightingfun}
	f(\lambda_{k},\lambda_{k'}) = \frac{i}{\lambda_{k} - \lambda_{k'}} \left(e^{-i\lambda_{k}} - e^{-i\lambda_{k'}}\right).
\end{equation}
We refer to this function as the first order weighting function. The above expressions are derived in %
 the Appendix, and are unsurprisingly in exact agreement with first-order time-dependent perturbation theory.
 
 Crucially, the Dyson operators depend on the size $\delta t$ of the time increment, but not the current time of the evolution~$t$. As a result, for a total evolution time $T = P\delta t$, where $P$ is an integer, the set of $P$ incremental evolution operators $\hat{U}(p\delta t, (p+1)\delta t)$ can be evaluated simultaneously. As will become clearer in the next section, this holds true to arbitrary order.

 \subsection{Evaluation to n-th order}
\label{sec:evaltoMthOrder}

To model quantum systems with sufficient accuracy, it is necessary to consider second- and higher-order terms in the Dyson series. This can be done following a similar approach as described above. %
Indeed, we introduce the $n$-th order Dyson operator in a similar fashion to \cref{eqn:Formal1stDysonOperator}
\begin{equation} \label{eqn:ShiftedfreqNthOrderDyson}
\begin{aligned}
	&\hat{S}^{(n)}(\bm{\omega}_n,\delta t)= \\
	&\left(\frac{-i\delta t}{2}\right)^n \sum_{\bm{k}_n} 
	f\left(\bm{\lambda}_n(\bm{k}_n)\delta t - \bm{c}(\bm{\omega}_n)\delta t \right)
	\langle k^{(n)}| \hat{X}| k^{(n-1)} \rangle \\ 
	&\times
	\langle k^{(n-1)} |\hat{X}\cdots|k^{(1)}\rangle 
	\langle k^{(1)} | \hat{X} | k^{(0)} \rangle 
	| k^{(n)} \rangle \langle k^{(0)} |.
\end{aligned}
\end{equation}
Here, each $\bm{k}_n = (k^{(0)},k^{(1)},\cdots,k^{(n)})$ is a set of indices which specify a set of $(n+1)$ eigenstates $\lbrace |k^{(m)} \rangle \rbrace$ of $\hat{H}_0$ with corresponding eigenvalues $\bm{\lambda}_n(\bm{k}_n)$, and we sum over all possible $\bm{k}_n$. These eigenvalues are written in vector form:
\begin{equation}
\label{eqn:VectorizedEigenvalues}
\bm{\lambda}_n (\bm{k}_n) \equiv \left(\lambda_{k^{(0)}},\cdots,\lambda_{k^{(n)}}\right), \quad \hat{H}_0 | k^{(m)} \rangle = \lambda_{k^{(m)}} | k^{(m)}\rangle.
\end{equation}
 The sum over $\bm{k}_n$ implies a summation over all sets of eigenstates which the dipole operator $\hat{X}$ couples 
 in \cref{eqn:ShiftedfreqNthOrderDyson}.
Additionally, we have introduced the cumulative vector %
\begin{equation}\label{eqn:Freqvector}
	\bm{c}(\bm{\omega}_n) = \left(\sum_{p=0}^{n-1} \bm{\omega}_n[n-p],\sum_{p=0}^{n-2} \bm{\omega}_n[n-p],\cdots,\bm{\omega}_n[n],0\right).
\end{equation}
The~$n$-th order weighting function $f(\bm{\lambda}_n)$ entering \cref{eqn:ShiftedfreqNthOrderDyson} can be obtained recursively using the relation (see \cref{app:DerivationNOrderSol})

\begin{equation} \label{eqn:Recursiveweightingfun}
f(\bm{\lambda}_n) =  i %
	\frac{f(\bm{g}(\bm{\lambda}_{n}))
	-f(\bm{g}^2(\bm{\lambda}_n)\cup \bm{\lambda}_n[n])}
{\bm{\lambda}_n[n-1] - \bm{\lambda}_n[n]}, %
\end{equation}
where $\bm{g}(\bm{v}_{n})$ returns %
$\bm{v}_n$ without its last %
element, $\bm{g}^2(\bm{v}_{n})= \bm{g}(\bm{g}(\bm{v}_{n}))$, and the notation~$\cup$ indicates appending an additional element to a vector %
such that
\begin{equation}
\bm{\lambda}_n = \bm{g}(\bm{\lambda}_{n}) \cup \bm{\lambda}_n[n].\\
\end{equation}
In the case of degenerate eigenvalues, we simply define \cref{eqn:Recursiveweightingfun} by taking the limit $\bm{\lambda}_n[n-1] \rightarrow \bm{\lambda}_n[n]$. %

Using \cref{eqn:OEdefn3} with the above results, we can now form the truncated Dyson series yielding an approximation to the evolution operator to $n$-th order in the perturbation
\begin{equation} \begin{aligned}
    \label{eqn:ParallelEv}
\hat{U}_p \approx \sum_{r=0}^n \hat{U}^{(n)}_p, \quad  \hat{U}^{(n)}_p \equiv \hat{U}^{(n)}(p\delta t,(p+1)\delta t).
\end{aligned}
\end{equation}
thus yielding the total evolution operator for time ~$T = P\delta t$ to the same order
\begin{equation} \label{eqn:MultAllEv}
\begin{aligned}
	\hat{U}(0,P\delta t) \approx \mathcal{T}\prod_{p=0}^P \hat{U}_p
\end{aligned}
\end{equation}
In this formalism, the time-ordering operator~$\mathcal{T}$ becomes a trivial operation  since we need only arrange the set of matrices $\lbrace\hat{U}_p\rbrace$ in ascending order from right to left. Using this method, we can thus calculate an arbitrary number of terms in the Dyson series, with each subsequent order increasing the accuracy of the approximation to the propagator operator $\hat{U}(0,T)$.

\subsection{Generalization to more complex drives}

So far, we have only considered a single, cosinusoidal drive. In practice, for applications such as DRAG~\cite{PhysRevLett.103.110501}, it is useful to consider more complicated drives of the form
\begin{equation}\label{eqn:sincosrealcomplexamp}
\begin{aligned}
	\hat{V}(t) &= \left[\Omega_x \cos(\omega t) + \Omega_y \sin(\omega t)\right]\hat{X},\\
 &= \left( \Omega e^{i\omega t} + \Omega^* e^{-i\omega t}\right)\hat{X},\\
\end{aligned}
\end{equation}
where~$\Omega = \Omega_x + i\Omega_y$ is the complex drive amplitude. This generalization requires only a minor modification to the result of \cref{eqn:OEdefn4} which now reads
\begin{equation} \label{eqn:FullUnitaryNthOrdersingledrive}
\hat{U}^{(n)}(t,t+\delta t) = \sum_{\bm{\omega}_n}\exp(i\sum_{p=1}^n\bm{\omega}_n[p] t) \Omega(\bm{\omega}_n) \hat{S}^{(n)}(\bm{\omega}_n,\delta t),
\end{equation}
and where we have introduced %
\begin{equation} \label{eqn:DriveFun1}
\begin{aligned}
\Omega(\bm{\omega}_n) &= \Omega ^{\mu(\bm{\omega}_n)} \Omega^{*(n-\mu(\bm{\omega}_n))}.
\end{aligned}
\end{equation}
We see from \cref{eqn:sincosrealcomplexamp} that $\Omega$ and $\Omega^*$ will appear according to the number of positive and negative frequency elements in the vector $\bm{\omega}$ respectively, leading to a simple expression for $\mu(\bm{\omega}_n)$ in $\Omega(\bm{\omega}_n)$:
\begin{equation}
\mu(\bm{\omega}_n)= (n + \sum_{p=1}^n\bm{\omega}_n[p])/(2\omega).
\end{equation}
As shown in the Appendix, this formalism can be further extended to an arbitrary number of drives of different frequencies, and acting on different system operators.

\subsection{Envelope Functions and Gaussian Filtering} \label{sec:Extraptomore}

Having established the necessary notation, we can now take into account control drives with time-dependent amplitudes $\Omega(t)$. To do so, we discretize the drive envelope of total time $T=N_p\Delta t$ into $N_p$ increments, called pixels, each of duration $\Delta t$. For a given pixel $i$, the drive is characterized by its complex amplitude $u_i$ such that the envelope can be expressed as \cite{KHANEJA2005296}
\begin{equation}
	\Omega(t) = \sum_{i=0}^{N_p-1}u_i\sqcap(i\Delta t,(i+1)\Delta t),
\end{equation}
where $\sqcap(t,t +\Delta t) = \Theta(t) - \Theta(t-\Delta t)$, with $\Theta(t)$ the Heaviside function.

\begin{figure}[t]
	\centering
	\includegraphics[width=1\columnwidth]{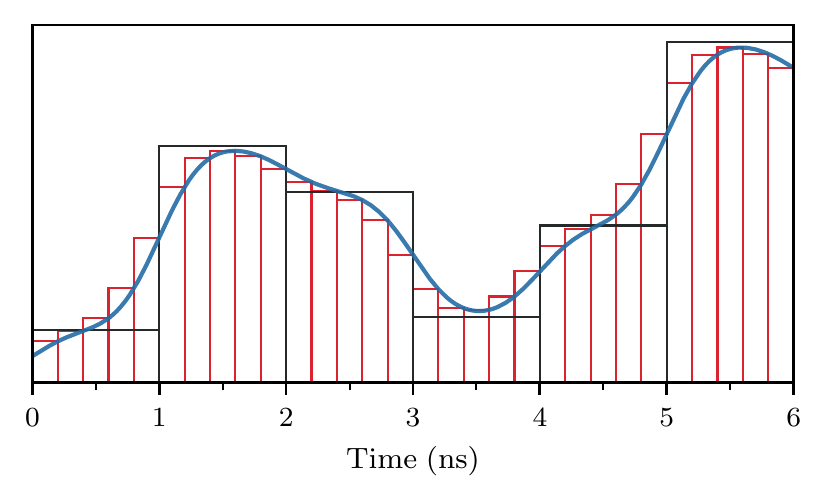}
	\caption{An example set of pulse amplitudes. The black bars indicate the chosen drive amplitudes $u_j$, where $\Delta t=1$~ns. The red bars indicate the subpixels, which provide an approximate interpolation of the true pulse delivered to the system (blue), with a bandwidth $\omega_0/2\pi = 851$ MHz.}
	\label{fig:Examplesubpixels}
\end{figure}

Additionally, following \textcite{FastTimeVaryingHam}, we take into account  the finite bandwidth of the control by applying a Gaussian filter on the discretized enveloped. To do so, each pixel is subdivided in $N_s$ subpixels of width $\delta t$ with $\Delta t = N_s \delta t$, see \cref{fig:Examplesubpixels}. The subpixel amplitudes $s_l$ are then defined as 
\begin{equation}
	s_l  = \sum_{j=0}^{N_s-1}T_{l,j}u_j,
\end{equation}
where the Gaussian filter matrix $T$ has elements
\begin{equation}
\begin{aligned}
	T_{l,j} &= \frac{1}{2}\left\lbrace \erf \left[ \omega_0 \left(\frac{l\delta t -j\Delta t}{2}\right)\right] \right.\\
	&\qquad\!\! - \left.\erf \left[ \omega_0 \left(\frac{l\delta t -(j+1)\Delta t}{2}\right)\right] \right\rbrace.
\end{aligned}
\end{equation}
Following \cref{eqn:FullUnitaryNthOrdersingledrive}, the $n$-th order evolution operator over the $l$-th subpixel takes the form
\begin{equation} \label{eqn:FullUnitaryNthOrdersingledrivesubpixel}
\begin{aligned}
\hat{U}^{(n)}_l = \sum_{\bm{\omega}_n}\exp(i\sum_{p=1}^n\bm{\omega}_n[p] l\delta t) \Omega_l(\bm{\omega}_n) \hat{S}^{(n)}(\bm{\omega}_n,\delta t),
\end{aligned}
\end{equation}
where the drive function in \cref{eqn:DriveFun1} picks up an additional subscript~$l$ to denote the~$l-$th subpixel
\begin{equation} \label{eqn:DriveFunctioneasy}
\begin{aligned}
\Omega_l(\bm{\omega}_n) &= s_l^{\mu(\bm{\omega}_n)} s_l^{* (n-\mu(\bm{\omega}_n))}.
\end{aligned}
\end{equation}

As can be seen in~\cref{fig:Examplesubpixels}, the subpixels (red) will often %
overestimate or underestimate the filtered pulse (blue) amplitude depending on its gradient, %
something which can become a leading contribution to the error in simulations.
In \cref{app:LinearInterpolation}, we generalize the amplitudes $s_l$ to have a linear time dependence to compensate for the change in amplitude of the continuous pulse across a single subpixel, providing a more accurate approximation to the filtered pulse.

\section{The \texttt{Dysolve} algorithm} \label{sec:DysolveAlg}

As already explained, the Dyson series operators $\hat{S}^{(n)}(\bm{\omega}_n,\delta t)$ for which we have expressions at arbitrary order $n$ are functions of the time increment $\delta t$ and, crucially, are independent of the total evolution time $T$. The \texttt{Dysolve} algorithm leverages this fact to parallelize the time evolution. 

The %
algorithm operates in two parts: a \textit{preparation} stage and a \textit{contraction} stage. In the preparation stage, the Dyson operators $\hat{S}^{(n)}(\bm{\omega}_n,\delta t)$ for a Hilbert space size $N$ are computed up to a chosen truncation order $n$, and arranged in a tensor.  This tensor has dimensions $(R\times N\times N)$, where $R$ is the total number of Dyson operators. In the case of a single drive without linear interpolation, $R = 2^{n+1} -1$, where $n$ is the order of the expansion.

Once the preparation stage is completed, it is in principle possible to consider arbitrary gate times and drive envelopes. Suppose we wish to simulate a time-evolution of length $T= P\delta t$, where $P$ is an integer. We first generate a $(P\times R)$ tensor whose elements are the envelopes and oscillatory terms $\exp(i\sum_{p=1}^n\bm{\omega}_n[p] l\delta t)\Omega_l(\bm{\omega}_n)$ in \cref{eqn:FullUnitaryNthOrdersingledrivesubpixel}. %
We then multiply these tensors to obtain a $(P \times N\times N )$ time-evolution tensor, where the $p$-th $(N\times N)$ matrix corresponds to a time-step operator $\hat{U}_p$. This multiplication constitutes the parallelized portion of the algorithm, with all $P$ time-evolution operators calculated simultaneously. %
As in \cref{eqn:MultAllEv}, we multiply all of the individual evolution operators in the time evolution tensor to obtain the final evolution operator $\hat{U}(0,T)$. This whole procedure forms the contraction step of our algorithm.

Below, we will refer to the computation of the evolution operator $\hat{U}(0,T)$ to $n$-order following the above approach as \texttt{Dysolve-n}.

\subsection{Benchmarking} \label{sec:Benchmarking}

\begin{figure*}
	\centering
\includegraphics[width=2\columnwidth]{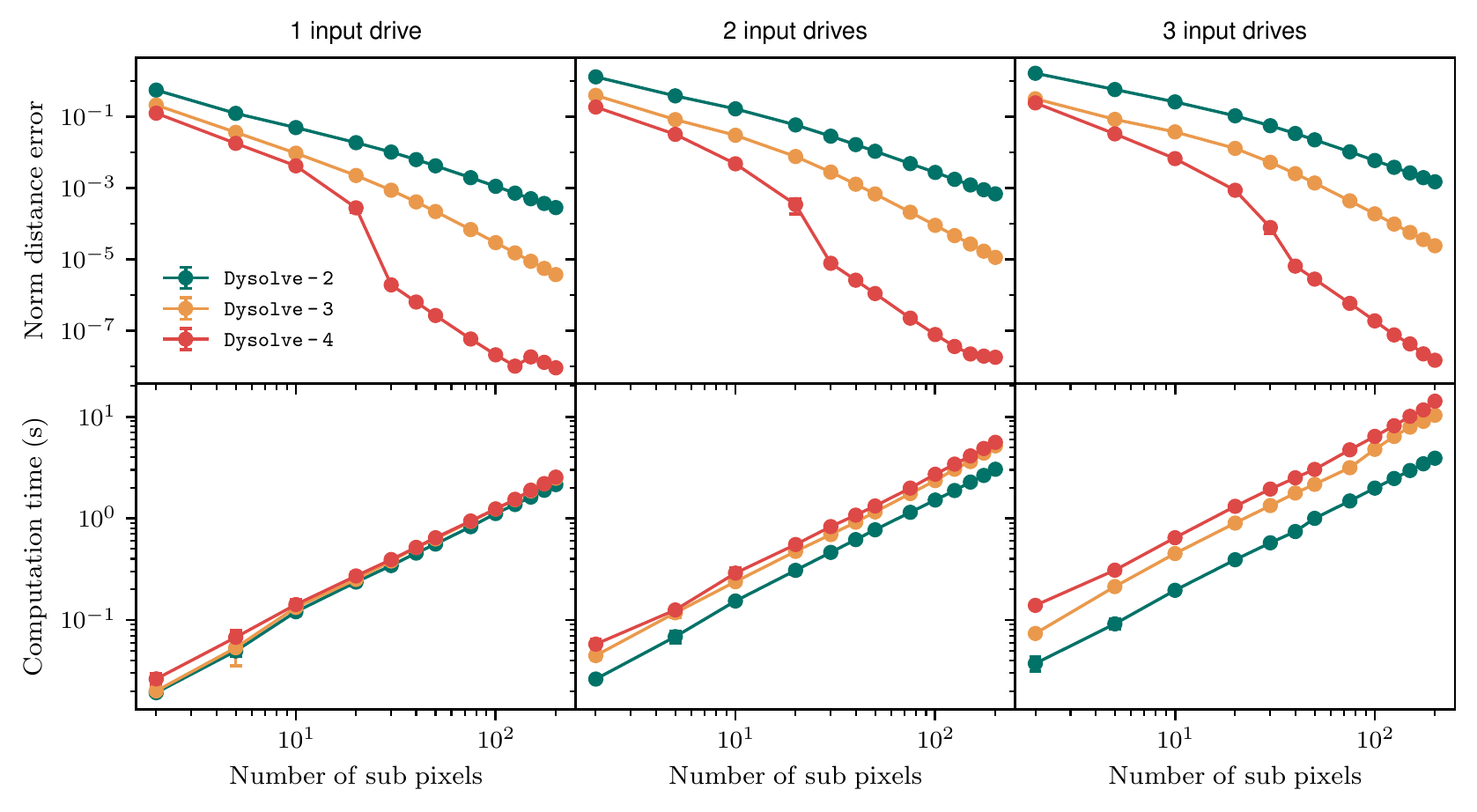}
	\caption{\texttt{Dysolve} benchmarks. First row: Frobenius norm distance from the propagator $\hat{U}_{\mathrm{ref}}(0,T)$ for the \texttt{Dysolve} algorithm for $T=500$ ns. Second row:  Contraction time of the \texttt{Dysolve} algorithm as a function of the subpixel number $N_s$. The slope of the data is precisely $1$, meaning that the computational time scales linearly with the number of subpixels. Preparation stage computation time: (35~ms, 1.9~s, 6.8~s) for (\texttt{Dysolve}-2, 3, 4) and 1 input drive, (0.35~s, 13~s, 87~s) for 2 input drives and  (0.78~s, 47~s, 408~s) for 3 input drives.}
	\label{fig: benchmark_6panels}
\end{figure*}

Before turning to examples of application of \texttt{Dysolve}, we first benchmark this algorithm. To do so, there are a number of factors to consider:
$i)$~the size of the Hilbert space,
$ii)$~the drive amplitude,
$iii)$~the number of independent drive channels, and
$iv)$~the shape of the envelope function.
To quantify the performance of our algorithm, we use two metrics. Given a number of subpixels $N_s$, we evaluate:
$1)$~the computation time, and
$2)$~the Frobenius norm distance between $\hat{U}(0,T)$, the propagator calculated under the \texttt{Dysolve} algorithm with a chosen number of subpixels,  and a reference $\hat{U}_{\textrm{ref}}(0,T)$, the same unitary calculated with very high %
precision. As discussed further in \cref{app:Benchmark}, we use \texttt{Dysolve}-4 with $10^4$ subpixels to compute $\hat{U}_{\textrm{ref}}(0,T)$, since \texttt{Dysolve} is able to reach precisions on the benchmark setup that are up to three orders of magnitude greater than QuTiP's \texttt{propagator}, a comparable dynamical solver \cite{Qutip2013}.

For our benchmarks, we use %
diagonal system Hamiltonians with an  Hilbert space size $N=25$. As a concrete example, we take the eigenvalues to be normally distributed about 7 GHz, the typical operating frequency of superconducting qubits \cite{Koch2007}. %
We consider between one and three input drive operators, %
at the frequency of the $\ket{0}\leftrightarrow \ket{1}$, $\ket{2}\leftrightarrow \ket{3}$ and $\ket{4}\leftrightarrow \ket{5}$ transitions with $\ket{k}$ the $k$-th lowest lying system eigenstate. The matrix elements of each operator corresponding to these transitions are set to $1$. To emulate an arbitrary drive operator with off-resonant terms, we populate %
$20\%$ of the remaining matrix elements of the drive operators %
with complex numbers normally distributed about 0, after which hermicity is enforced by the addition of the complex conjugate. The envelope function associated to each drive operator is centered around an amplitude such that the duration of the simulation is equivalent, in the absence of the other drives, to a total of 20 Rabi oscillations. %
In the context of superconducting qubits with their microwave drives, this corresponds to the simulation a 500~ns evolution  with a drive amplitude of 40~MHz. We take the pixel amplitudes $u_j$ %
of the envelope functions to be normally distributed about 40~MHz with a standard deviation of 1~MHz. To reduce statistical fluctuations, the results presented in \cref{fig: benchmark_6panels} are averaged over 30 simulations, each with different random envelope functions and system eigenvalues.

 The numerical simulations reported in \cref{fig: benchmark_6panels} are performed on a 2.8~GHz, Intel Xeon Gold 6242 Processor (16 cores/32 threads) using \texttt{python}.
Since the preparation stage only needs to be performed once for a particular system Hamiltonian, %
the reported computation time accounts only for the contraction stage of the \texttt{Dysolve} algorithm. We report the preparation times in the figure caption for reference. The top three panels of \cref{fig: benchmark_6panels} present the norm distance between \texttt{Dysolve-n} for \texttt{n} =  \texttt{2}, \texttt{3} and \texttt{4} as a function of the number of subpixels, and for increasing number of input drives. Even with large drive amplitudes and long evolution times, we obtain an excellent approximation to the final unitary operation with relatively few subpixels. For example, using a fourth order Dyson expansion with $40$ subpixels yields a norm distance error of less than $10^{-5}$ in all cases. Importantly, as shown in the bottom panels of \cref{fig: benchmark_6panels}, the computation time needed to reach this level of accuracy is only on the order of a few seconds (less than 3~s). %
In comparison, QuTiP's \texttt{propagator} function takes about $16$ minutes to perform the calculation to an equivalent accuracy for a single input drive. Such a significant speedup proves to be particularly useful when, as dicussed in section~\cref{sec:GRAPE}, the contraction stage of the algorithm needs to be repeated many times in an optimization loop.
Additional benchmarking results are provided in \cref{app:Benchmark}.

\section{Application to Optimal Control} \label{sec:GRAPE}

Optimal control is an essential tool in the development of high-fidelity gates for quantum computation. 
Most optimization algorithms, such as GRAPE~\cite{KHANEJA2005296}, rely on calculating the gradient of the gate fidelity with respect to the drive amplitude at each pixel. With most approaches, this requires recalculating the evolution operator at each time interval, something which can be as expensive as the original calculation of the unitaries. Moreover, this calculation is generally performed with the presumption that the Hamiltonian is time independent over the duration of a subpixel, thus invoking a form of the RWA. In our expansion, such an assumption is not required. %

Here, we show how \texttt{Dysolve} can be applied to optimizing control pulse envelopes to maximize gate fidelity. More precisely, we consider optimizing the fidelity of an evolution $\hat{U}(T)$ with respect to a target gate unitary $\hat{U}_{\textrm{target}}$. In general, $\hat{U}(T)$ acts on the full Hilbert space of the system while $\hat{U}_{\textrm{target}}$ is defined on its computational subspace. The objective is thus to maximize the performance function \cite{FastTimeVaryingHam}
\begin{equation}\label{eqn:fidelity}
	\Phi = \frac{1}{d^2}\abs{\Tr( \hat{U}^\dagger_{\textrm{target}}\hat{U}(T)\hat{\mathcal{P}} )}^2,
\end{equation}
where~$d$ is the dimension of the computational subspace ($d=2$ for a single qubit), and $\hat{\mathcal{P}}$  is the projector on that subspace. Although our approach can in principle deal with an arbitrary number of drives, for simplicity here we consider control of a single set of complex drive amplitudes $u_j$. 

We use a GRAPE-like approach to maximize the gate fidelity which requires the gradient of the cost function $\Phi$ with respect to the drive amplitude at each pixel~$u_j$ and subpixel $s_l$~\cite{FastTimeVaryingHam}
\begin{equation} \label{eqn:Fidelityderivative}
\begin{aligned}
	&\frac{\partial \Phi}{\partial u_{j}} = \sum_{l=0}^{M-1}T_{l,j} \frac{\partial \Phi}{\partial s_{l}},\\
	&\frac{\partial \Phi}{\partial s_{l}} = \frac{1}{d^2}2 \Re{\Tr \left[\hat{U}_\textrm{target}^\dagger \frac{\partial \hat{U}}{\partial s_{l}} \mathcal{P} \right] \Tr \left[\hat{U}_\textrm{target} \hat{U}^\dagger \mathcal{P} \right] }.
	\end{aligned}
\end{equation}
Within the framework of the \texttt{Dysolve} algorithm, it is simple to evaluate the operator $(\partial \hat{U}/\partial s_l)$. Indeed, using %
\cref{eqn:MultAllEv}, we find that for the $n$-th order Dyson expansion, the derivatives of the unitaries take the form %
\begin{equation}
\begin{aligned}
    &\frac{\partial \hat{U}}{\partial s_l} =\left( \prod_{m=l+1}^M \hat{U}_m\right) \frac{\partial \hat{U}_l}{\partial s_l} \left(\prod_{p=0}^{l-1} \hat{U}_p\right)  ,\\
    &\frac{\partial \hat{U}_l}{\partial s_l} =\!\!\sum_{m=0}^n \sum_{\bm{\omega}_m}\exp(\!\!i\sum_{p=1}^m\bm{\omega}_m[p]l\delta t\!\!) \frac{\partial \Omega_l(\bm{\omega}_m)}{\partial s_{l}} \hat{S}^{(m)}(\bm{\omega}_m,\delta t),\\
     &\frac{\partial \Omega_l(\bm{\omega}_m)}{\partial s_{l}} = \mu(\bm{\omega}_m) s_{l}^{(\mu(\bm{\omega}_m)-1)}s_{l}^{* (n-\mu(\bm{\omega}_m))}.
\end{aligned}
\end{equation}
Importantly, note that the Dyson operators $\hat{S}^{(n)}(\bm{\omega}_n,\delta t)$ remain unchanged by the derivative. As such we only need to perform the preparation stage of the \texttt{Dysolve} algorithm once, with the calculation of $\partial\hat{U}/\partial s_l$. Thus, the optimization iterations only require the contraction stage computation to be performed. Further, these derivatives are exact. Consequently, the effects of off-resonant and counter-rotating terms are accounted for in the derivatives. This is the strength of the \texttt{Dysolve} algorithm for optimization. 

Recall from \cref{eqn:sincosrealcomplexamp} that the real and imaginary components of the drive amplitudes are associated with the magnitudes of the cosine and sine quadratures, respectively. Thus, to calculate the relevant amplitude derivatives for the cosine and sine drive envelopes, one simply calculates the appropriate sum or difference of the derivatives in \cref{eqn:Fidelityderivative}:
\begin{equation} \label{eqn:Derivswrtrealimag}
\begin{aligned}
 \frac{\partial \Phi }{\partial u_{j,x}} = \frac{1}{2}\left(\frac{\partial \Phi }{\partial u_{j}} + \frac{\partial \Phi }{\partial u^*_{j}} \right),\\
  \frac{\partial \Phi }{\partial u_{j,y}} = \frac{-i}{2}\left(\frac{\partial \Phi }{\partial u_{j}} - \frac{\partial \Phi }{\partial u^*_{j}} \right),
\end{aligned}
\end{equation}
where $u_j= u_{j,x} + i u_{j,y}$. To perform the GRAPE algorithm, the set of drive amplitudes are simply updated by taking steps in the direction of the gradient of the fidelity~\cite{KHANEJA2005296}
\begin{equation}
    \begin{aligned}
    u_{j,(x,y)} \rightarrow u_{j,(x,y)} + \epsilon \frac{\partial \Phi}{\partial u_{j,(x,y)}},
    \end{aligned}
\end{equation}
where~$\epsilon$ is a small, positive number which need not be fixed during the optimization process. For example, one could use a backtracking line search to maximize the gain in fidelity \cite{armijo1966,Backtracklinesearch}. In the examples presented below, we simply use a sufficiently small $\epsilon$ for the updating process to be stable.

\subsection{Example: Cross-Resonance Gate for Coupled Transmons}

\begin{figure}[t] 
	\centering
	\includegraphics[width=0.75\columnwidth]{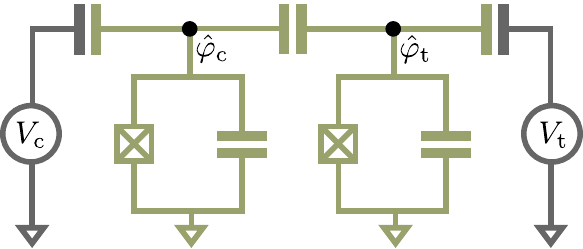}
	\caption{Superconducting circuit for cross-resonance gates between transmon qubits. The leftmost transmon, of frequency~$\omega_\mathrm{c}/2\pi$, plays the role of the control qubit and is strongly driven by the voltage source~$V_\mathrm{c}$ at the frequency~$\omega_\mathrm{t}/2\pi$ of the target qubit (rightmost transmon). The target qubit is also driven by~$V_\mathrm{t}$ using a relatively weak tone that serves to give additional control. $\hat{\varphi}_\mathrm{c}$ and~$\hat{\varphi}_\mathrm{t}$ correspond to the phase degree of freedom associated with the control and target qubits, respectively.}
	\label{fig: cr_gate}
\end{figure}

As an example of application of our implementation of the GRAPE algorithm with \texttt{Dysolve}, we present results of the optimization of a two-qubit gate. To this end, we consider the system of two fixed-frequency transmon qubits, illustrated in \cref{fig: cr_gate}. Each qubit is described by a Hamiltonian of the form~\cite{Koch2007}
\begin{equation}
    \hat{H}_j = 4E_{Cj}\hat{n}_j^2 - E_{Jj}\cos\hat{\varphi}_j,
\end{equation}
where $\hat{n}_j$ and $\hat{\varphi}_j$ with $j=c,t$ are the conjugate charge and phase operators of the transmon, while $E_{Cj}$ and $E_{Jj}$ are the charging and Josephson energies. %
More precisely, we consider the  cross-resonance gate, which performs an $X$ rotation on a target qubit conditional on the state of a control qubit which is driven at the target qubit's frequency \cite{Rigetti2010,chow2011simple}. %
Optimal control has been applied to this gate before in Refs.~\cite{allen2017optimal,kirchhoff2018optimized}. However, these approaches have made use of simplified models for the device and of the rotating wave approximation.Taking advantage of the \texttt{Dysolve} algorithm, our approach generalizes previous work on the cross-resonance gate by considering the full circuit Hamiltonian and including all counter-rotating terms.

Following the standard circuit-quantization procedure~\cite{devoret1995quantum}, the two-transmon Hamiltonian can be put in the form $\hat{H}(t) = \hat H_0 + \hat V(t)$ with 
\begin{equation}
    \begin{aligned}
    \hat{H}_0 &= \hat{H}_{c} + \hat{H}_{t} + \hbar g \hat{n}_{c}\hat{n}_{t}, \\ 
    \hat{V}(t) &= \Omega_c(t)\hat{n}_c + \Omega_t(t)\hat{n}_t,
    \end{aligned}
\end{equation}
where $\hat{H}_{c},\hat{H}_{t}$ are the Hamiltonian of the control and target qubits, respectively, and $\hbar g\hat{n}_c\hat{n}_t$ results from the capacitive interaction between the qubits. %
The amplitudes $\Omega_c(t),\Omega_t(t)$ are the time-dependent drives
\begin{equation}\label{eq:TransmonDrives}
    \begin{aligned}
    \Omega_c(t) &= \left[ \Omega_{cx}(t)\sin(\omega_t t) + \Omega_{cy}(t)\cos(\omega_c t)\right], \\
    \Omega_t(t) &= \left[ \Omega_{tx}(t)\sin(\omega_t t) + \Omega_{ty}(t)\cos(\omega_t t)\right], \\
    \end{aligned}
\end{equation}
with $\lbrace \Omega_{cx},\Omega_{cy},\Omega_{tx},\Omega_{ty}\rbrace$ the drive envelope functions to be optimized by GRAPE. While activating the cross-resonance gate only requires driving the control qubit at the frequency of the target, the additional control over the target qubit $\hat{\Omega}_t(t)$ is useful to eliminate single qubit rotations and obtain higher fidelities to the target unitary~\cite{sheldon2016procedure,patterson2019calibration}.

\begin{figure}[t]
	\centering
	\includegraphics[width=1\columnwidth]{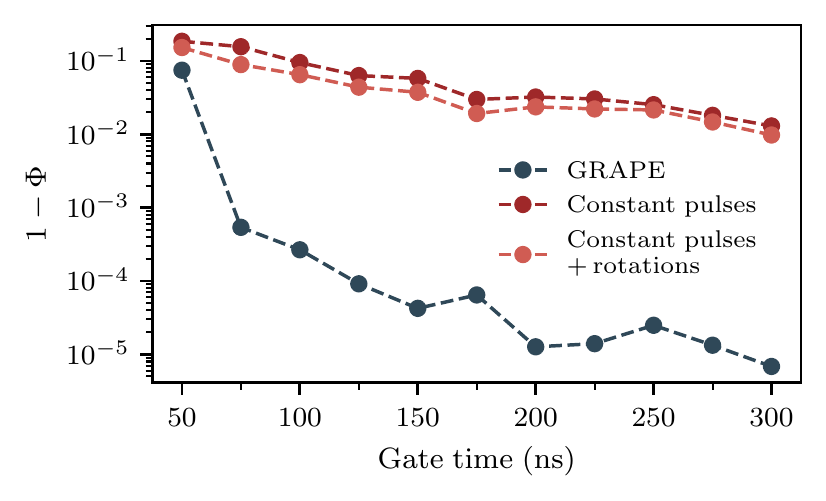}
	\caption{Infidelity of the CR gate as a function of the chosen gate time when the gate is operated at the qubit-qubit detuning $\Delta/2\pi = (\omega_c - \omega_t)/2\pi = 210$ MHz. The red curve indicates the best obtainable infidelity from a constant amplitude pulse, with the orange curve reporting the infidelity up to arbitrary single qubit rotations on both qubits, applied before and after the cross-resonance gate channel. The blue curve reports the infidelity when optimized under the GRAPE algorithm until convergence.}
	\label{fig: GRAPE-vs-Flat}
\end{figure}

We choose to operate this gate with control and target frequencies of $\omega_c/2\pi = 5.1$~GHz, $\omega_t/2\pi = 4.9$~GHz, and anharmonicities $\alpha_c/2\pi = -355$~MHz and $\alpha_t/2\pi = -352$~MHz, respectively. Moreover, the qubit-qubit coupling is set to $g/2\pi = 4.29$ MHz and the target unitary is taken to be $ZX_{90}$ gate~\cite{chow2012universal,corcoles2013process}. \Cref{fig: GRAPE-vs-Flat} shows the infidelity of the cross-resonance gate as a function of the gate time. We find that for a flat pulse a gate fidelity of $98.5\%$ in a $300$~ns gate time is possible (red symbols), while values approaching 99\% can be reached when correcting for additional single-qubit rotations (orange symbols). %
In contrast, GRAPE reaches significantly higher fidelities, going beyond $99.99\%$ in a shorter gate time, even when accounting for all off-resonant and counter rotating terms (blue symbols).

\begin{figure}[t]
	\centering\includegraphics[width=1\columnwidth]{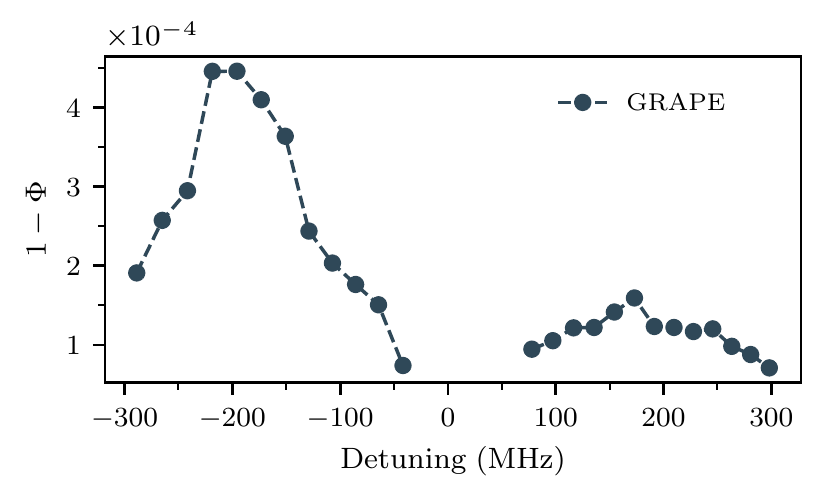}
	\caption{Infidelity of the CR gate as a function of the detuning of the qubits  when optimized under GRAPE for 5000 iterations.  The control qubit frequency is fixed to $\omega_c/2\pi=5.10$~GHz.}
	\label{fig: Detunings}
\end{figure}

To further demonstrate that GRAPE under $\texttt{Dysolve}$ can successfully optimize the cross-resonance gate more generally, \cref{fig: Detunings} shows the fidelity of an optimized 300-ns gate as a function of the qubit-qubit detuning $\Delta/2\pi = (\omega_c - \omega_t)/2\pi$. In the numerical simulations, this detuning is varied by changing the target qubit frequency  
whilst keeping the qubit-qubit coupling and the control-qubit frequency fixed. We note that the performance of the gate at positive detunings is approximately in line with those predicted by Schrieffer-Wolff perturbation theory for an echoed-cross resonance gate performed on a similar device \cite{Malekakhlagh2020}. The results are excluded at~$\Delta = 0,\pm \alpha$, as the qubits strongly hybridized due to resonances. The slight variations in the GRAPE fidelity in the wide detuning range are partially attributable to variations in higher-order two-qubit coupling amplitudes across, alongside the performance of gradient ascent for different optimization landscapes. This also constitutes evidence for the robustness of our algorithm, as excellent fidelities are obtained in a broad range of parameters. 

\section{Future Work and Conclusion}

We have demonstrated that the \texttt{Dysolve} algorithm provides a means to quickly and accurately simulate driven systems whilst accounting for all of the effects of the counter-rotating and off-resonant terms. Analysis of ultrafast quantum gates and the quantum speed limit~\cite{Shao_2020}, where counter-rotating effects are particularly important, would also be possible with our algorithm. Additionally, this method trivializes the calculation of the gradient, allowing for rapid optimization without the need for additional approximations, and can be modified to include dissipative effects. Indeed, a simple extension of the optimization scheme would allow for optimization of lossy quantum systems specifically, through an open GRAPE-like scheme \cite{PhysRevA.99.052327}, or the simulation of larger lossy systems through the use of trajectories. We also note that as the expressions depend explicitly on the drive amplitudes, second derivatives and the Hessian matrix can also be calculated exactly, opening the door to second order optimizers such as Newton's method.

We anticipate that future iterations of the \texttt{Dysolve} algorithm improve the efficiency of both the preparation and contraction stages of the algorithm. Given the parallelized nature of this solver, we also envision a direct extension to graphics processing units (GPU's), which could allow for fast simulation of significantly larger systems. We leave this for future work.

At the time of publication, the authors were made aware of a recent paper~\cite{kalev2020integralfree} which uses a different approach to obtain a result similar to that which we provide in \cref{sec:GenericSinusoidal}. Our derived weighting functions are equivalent to divided differences~\cite{Divided_Differences2005} utilised in their method.

\begin{acknowledgments}
	We thank Catherine Leroux for discussions and \'Eric Gigu\`ere for assistance with numerical calculations.
	This work was undertaken thanks in part to funding from NSERC, the Canada First Research Excellence Fund, and the U.S. Army Research Office Grant No. W911NF-18-1-0411.
\end{acknowledgments}

\addcontentsline{toc}{section}{References}
\bibliography{biblx}{}

\appendix

\onecolumngrid

\section{Derivation of n-th order weighting functions}
\label{app:DerivationNOrderSol}

In this Appendix, we derive the $n$-th order weighting function and the form of the Dyson series operators. We begin by defining a frequency-dependent set of path operators
\begin{align}\label{eqn:NewPathOps}
\hat{R}_{[m_n,\ldots,m_0]}(\bm{v}_n,\delta t)
&= (-i)^n
\int_{0}^{\delta t}dt_M\cdots\int_{0}^{t_2}dt_1
(-i(\hat{H}_0 - \bm{v}_n[n]))^{m_n}\hat{X}\cdots (-i(\hat{H}_0 - \bm{v}_n[1]))^{m_1}\hat{X}(-i(\hat{H}_0 -i\bm{v}_n[0]))^{m_0},
\end{align}
where, for now, $\bm{v}_n$ is an arbitrary vector and %
$M = \sum_{i=0}^n m_i + n$. Since there is no explicit time dependence in \cref{eqn:NewPathOps}, we can evaluate its %
$M$ integrals simultaneously,
\begin{equation}
	\begin{split}
		\hat{R}_{[m_n,\ldots,m_0]}(\bm{v}_n,\delta t)
		&=\frac{(-i)^n(-i(\hat{H}_0- \bm{v}_n[n]))^{m_n}\hat{X}\cdots (-i(\hat{H}_0- \bm{v}_n[1]))^{m_1}\hat{X}(-i(\hat{H}_0- \bm{v}_n[0]))^{m_0} (\delta t)^M}{M!}\\ 
		&=
		(-i\delta t)^n
		\frac{(-i\delta t(\hat{H}_0-\bm{v}_n[n]))^{m_n}\hat{X}\cdots(-i\delta t(\hat{H}_0- \bm{v}_n[0]))^{m_0}}
		{M!}.
\end{split}
\end{equation}
Inserting identity operators $I_N = \sum_k |k\rangle \langle k|$, with $ \hat{H}_0|k\rangle = \lambda_k |k\rangle$ and $N$ the size of the system's Hilbert space, between groups of $\hat{H}_0$ and each $\hat{X}$ operator leads to
\begin{equation} \label{eqn:PathOperatorExpanded}
	\begin{split}
		\hat{R}_{[m_n,\ldots,m_0]}(\bm{v}_n,\delta t)
&=
	(-i\delta t)^n \sum_{\bm{k}_n} \langle k^{(n)}| \hat{X}| k^{(n-1)} \rangle \langle k^{(n-1)} |\hat{X}...|k^{(1)}\rangle \langle k^{(1)} | \hat{X} | k^{(0)} \rangle | k^{(n)} \rangle \langle k^{(0)} | 
	\\
	&\quad
	\times\frac{(-i(\bm{\lambda}_n(\bm{k}_n) - \bm{v}_n)[0]\,\delta t)^{m_0}\cdots
	(-i(\bm{\lambda}_n(\bm{k}_n) - \bm{v}_n)[n]\,\delta t)^{m_n}}
	{M!}.
	\end{split}
\end{equation}
To obtain the Dyson operator, we sum \cref{eqn:PathOperatorExpanded} over all $\bm{m}\in\mathbf{Z}_+^{n+1}$. To this end, we define the $n$-th order weighting function $f(\bm{\lambda}_n)$ as
\begin{equation}
	\begin{split}
		f(\bm{\lambda}_n)
	&=
	\sum_{m_0,\ldots,m_n}\frac{(-i\bm{\lambda}_n[0])^{m_0}\cdots(-i\bm{\lambda}_n[n])^{m_n}}
	{\big(\sum_{z=0}^n(m_z)+n\big)!}
	\\
	&=
	\sum_{m_0,\ldots,m_{n-2}}
	(-i\bm{\lambda}_n[0])^{m_0}\cdots(-i\bm{\lambda}_n[n-2])^{m_{n-2}}
	\sum_{P=0}^\infty\sum_{p=0}^P
	\frac{(-i\bm{\lambda}_n[n-1])^{P-p}(-i\bm{\lambda}_n[n])^p}
	{\big(\sum_{n=0}^{n-2}(m_n)+P-p+p+n\big)!}
	\\
	&=
	\sum_{m_0,\ldots,m_{n-2},P}
	\frac{(-i\bm{\lambda}_n[0])^{m_0}\cdots(-i\bm{\lambda}_n[n-2])^{m_{n-2}}\big((-i\bm{\lambda}_n[n-1])^{P+1}-(-i\bm{\lambda}_n[n])^{P+1}\big)}
	{\big(\sum_{n=0}^{n-2}(m_n)+P+1+n-1\big)!(-i\bm{\lambda}_n[n-1]+i\bm{\lambda}_n[n])}
	\\
	&=
	i\left[\frac{f(\bm{g}(\bm{\lambda}_{n}))
		-f(\bm{g}^2(\bm{\lambda}_n) \cup \bm{\lambda}_n[n])}
	{\bm{\lambda}_n[n-1]-\bm{\lambda}_n[n]}\right],
	\end{split}
\label{eqn:derivationofnthweightfunction}
\end{equation}
where $f(\bm{\lambda}_0) = \exp(-i\bm{\lambda}_0[0])$. This gives us the form of $\hat{R}^{(n)}(\bm{v}_n,0,\delta t)$:
\begin{align}
\hat{R}^{(n)}(\bm{v}_n,\delta t)
&=
	\left(\frac{-i\delta t}{2}\right)^n \sum_{\bm{k}_n} \langle k^{(n)}| \hat{X}| k^{(n-1)} \rangle \langle k^{(n-1)} |\hat{X}...|k^{(1)}\rangle \langle k^{(1)} | \hat{X} | k^{(0)} \rangle | k^{(n)} \rangle \langle k^{(0)} | \times
	f((\bm{\lambda}_n(\bm{k}_n) - \bm{v}_n)\,\delta t).
\end{align}
We then note that
\begin{align}
	e^{i\omega t_0}f(\bm{\lambda}_n) =  f(\bm{\lambda}_n - \omega t_0).
	\label{eqn:Shiftingpropertyweightingfunction}
\end{align}
To prove this, first note that this is trivially true for %
$f(\bm{\lambda}_0)$. We then make the inductive step, assuming \cref{eqn:Shiftingpropertyweightingfunction} to be true for $n-1$. Then, %
\begin{equation}
	\begin{split}
		e^{i\omega t_0}f(\bm{\lambda}_n) &=  i\left[\frac{e^{i\omega t_0} f(\bm{g}(\bm{\lambda}_{n}))
		-e^{i\omega t_0} f(\bm{g}^2(\bm{\lambda}_n)\cup \bm{\lambda}_n[n])}
	{\bm{\lambda}_n[n-1] - \bm{\lambda}_n[n]} \right]\\
	&=  i\left[\frac{f(\bm{g}(\bm{\lambda}_{n})-\omega_0 t)
		 f(\bm{g}^2(\bm{\lambda}_n)\cup \bm{\lambda}_n[n] - \omega_0 t)}
	{\bm{ (\lambda}_n[n-1] - \omega_0 t) - (\bm{\lambda}_n[n] - \omega_0 t)} \right]\\
	&= f(\bm{\lambda}_n - \omega_0 t).
	\end{split}
\end{equation}
By linearity, this implies that $e^{ia\delta t}\hat{R}^{(n)}(\bm{v}_n,\delta t) = \hat{R}^{(n)}(\bm{v}_n + a,\delta t)$.\\

We now consider the effect of a set of oscillatory terms. To this end, we first note that the Dyson path operator in \cref{eqn:PathOperatorNthOrder}, as well as the Dyson operators themselves, can be defined recursively
\begin{align}
    \hat{S}^{(n)}_{[m_n,\ldots,m_0]}(\bm{\omega}_n,\delta t)&=(-i\hat{H}_0)^{m_n}\hat{X}\int_0^{\delta t}dt_M\cdots\int_0^{t_{M-m_n+1}}\hspace{-3em}dt_{M-m_n} e^{i\bm{\omega}_n[n]t_{M-m_{n}}}\hat{S}^{(n-1)}_{[m_{n-1},\ldots,m_0]}(\bm{g}(\bm{\omega}_{n}),t_{M-m_n})\,,\\
    \hat{S}^{(n)}(\bm{\omega}_n,\delta t)&=\frac{1}{2}\sum_{\mathclap{m_n\in\mathbf{Z}_+}}(-i\hat{H}_0)^{m_n}\hat{X}\int_0^{\delta t}dt_M\cdots\int_0^{t_{M-m_n+1}}\hspace{-3em}dt_{M-m_n} e^{i\bm{\omega}_n[n]t_{M-m_{n}}}\hat{S}^{(n-1)}(\bm{g}(\bm{\omega}_{n}),t_{M-m_n})\,,\\
    \hat{R}^{(n)}(\bm{v}_n,\delta t)&=\frac{1}{2}\sum_{\mathclap{m_{n}\in\mathbf{Z}_+}}(-i\hat{H}_0 - i\bm{v}_n[n])^{m_n}\hat{X}\int_0^{\delta t}dt_M\cdots\int_0^{t_{M-m_n+1}}\hspace{-3em}dt_{M-m_n} \hat{R}^{(n-1)}(\bm{g}(\bm{v}_{n}),t_{M-m_n})\,,
\end{align}
We now wish to demonstrate that $\hat{R}^{(n)}(\bm{c}(\bm{\omega}_n),\delta t) = \hat{S}^{(n)}(\bm{\omega}_n,\delta t)$, where $\bm{c}(\bm{\omega}_n)$ is the cumulative vector introduced in \cref{eqn:Freqvector}. To first order,
\begin{equation}
	\begin{split}
		\hat{S}^{(1)}(\bm{\omega}_1,\delta t)&=\frac{1}{2}\sum_{\mathclap{m_1\in\mathbf{Z}_+}}(-i\hat{H}_0)^{m_1}\hat{X}\int_0^{\delta t}dt_M\cdots\int_0^{t_{M-m_1+1}}\hspace{-3em}dt_{M-m_1} e^{i\bm{\omega}_1[1]t_{M-m_{1}}}\hat{S}^{(0)}(\bm{0},t_{M-m_1})\\
    &=\frac{1}{2}\sum_{\mathclap{m_1\in\mathbf{Z}_+}}(-i\hat{H}_0)^{m_1}\hat{X}\int_0^{\delta t}dt_M\cdots\int_0^{t_{M-m_1+1}}\hspace{-3em}dt_{M-m_1} \left[ \sum_{i} |i\rangle \langle i |  e^{i\bm{\omega}_1[1]t_{M-m_{1}}} f(\lambda_i t_{M-m_{1}}) \right]\\
    &=\frac{1}{2}\sum_{\mathclap{m_1\in\mathbf{Z}_+}}(-i\hat{H}_0)^{m_1}\hat{X}\int_0^{\delta t}dt_M\cdots\int_0^{t_{M-m_1+1}}\hspace{-3em}dt_{M-m_1} \left[ \sum_{i} |i\rangle \langle i | f((\lambda_i - \bm{\omega}_1[1]) t_{M-m_{1}})\right],\\
    &=\frac{1}{2}\sum_{\mathclap{m_1\in\mathbf{Z}_+}}(-i\hat{H}_0)^{m_1}\hat{X}\int_0^{\delta t}dt_M\cdots\int_0^{t_{M-m_1+1}}\hspace{-3em}dt_{M-m_1}\hat{R}^{(0)}([\bm{\omega}_1[1]],t_{M-m_1}),\\
    &=\frac{1}{2}\sum_{\mathclap{m_1\in\mathbf{Z}_+}}(-i(\hat{H}_0 -0))^{m_1}\hat{X}\int_0^{\delta t}dt_M\cdots\int_0^{t_{M-m_1+1}}\hspace{-3em}dt_{M-m_1}\hat{R}^{(0)}(\bm{g}([\bm{\omega}_1[1],0]),t_{M-m_1}),\\
    &=\hat{R}^{(1)}([\bm{\omega}_1[1],0],\delta t)\\
    &=\hat{R}^{(1)}(\bm{c}(\bm{\omega}_{1}),\delta t).
	\end{split}
\end{equation}
We again make the inductive step with the assumption that $\hat{R}^{(n-1)}(\bm{c}(\bm{\omega}_n),\delta t) = \hat{S}^{(n-1)}(\bm{\omega}_n,\delta t)$. To proceed, we first show the following result:
\begin{equation}
	\begin{split}
		\bm{c}(\bm{g}(\bm{v}_n)) &= \left(\left[\sum_{p=1}^{n-1} \bm{v}_n[n-p]\right],\left[\sum_{p=1}^{n-2} \bm{v}_n[n-p]\right],\cdots,\bm{v}_n[n-1],0\right).\\
    	&= \left(\left[\sum_{p=0}^{n-1} \bm{v}_n[n-p]\right],\left[\sum_{p=0}^{n-2} \bm{v}_n[n-p]\right],\cdots,\bm{v}_n[n-1] + \bm{v}_n[n],\bm{v}_n[n]\right) - \bm{v}_n[n]\\
        &= \bm{g}\left(\left[\sum_{p=0}^{n-1} \bm{v}_n[n-p]\right],\left[\sum_{p=0}^{n-2} \bm{v}_n[n-p]\right],\cdots,\bm{v}_n[n-1] + \bm{v}_n[n],\bm{v}_n[n],0\right) - \bm{v}_n[n]\\
    	&= \bm{g}(\bm{c}(\bm{v}_n))- \bm{v}_n[n]\\
	\end{split}
\end{equation}
Then,
\begin{equation}
	\begin{split}
		&\hat{S}^{(n)}(\bm{\omega}_{n},\delta t)\\
&=\frac{1}{2}\sum_{\mathclap{m_n\in\mathbf{Z}_+}}(-i\hat{H}_0)^{m_n}\hat{X}\int_0^{\delta t}dt_M\cdots\int_0^{t_{M-m_n+1}}\hspace{-3em}dt_{M-m_n} e^{i\bm{\omega}_n[n]t_{M-m_{n}}}\hat{S}^{(n-1)}(\bm{g}(\bm{\omega}_{n}),t_{M-m_n})\\
&=\frac{1}{2}\sum_{\mathclap{m_n\in\mathbf{Z}_+}}(-i\hat{H}_0)^{m_n}\hat{X}\int_0^{\delta t}dt_M\cdots\int_0^{t_{M-m_n+1}}\hspace{-3em}dt_{M-m_n} e^{i\bm{\omega}_n[n]t_{M-m_{n}}}\hat{R}^{(n-1)}(\bm{c}(\bm{g}(\bm{\omega}_{n})),t_{M-m_n})\\
&=\frac{1}{2}\sum_{\mathclap{m_n\in\mathbf{Z}_+}}(-i\hat{H}_0)^{m_n}\hat{X}\int_0^{\delta t}dt_M\cdots\int_0^{t_{M-m_n+1}}\hspace{-3em}dt_{M-m_n} \hat{R}^{(n-1)}(\bm{c}(\bm{g}(\bm{\omega}_{n})) + \bm{\omega}_n[n],t_{M-m_n})\\
&=\frac{1}{2}\sum_{\mathclap{m_n\in\mathbf{Z}_+}}(-i\hat{H}_0 - i0)^{m_n}\hat{X}\int_0^{\delta t}dt_M\cdots\int_0^{t_{M-m_n+1}}\hspace{-3em}dt_{M-m_n} \hat{R}^{(n-1)}(\bm{g}(\bm{c}(\bm{\omega}_{n})),t_{M-m_n})\\
&=\hat{R}^{(n)}(\bm{c}(\bm{\omega}_{n}),\delta t).\\
	\end{split}
\end{equation}
finishing the induction and giving the final desired form in \cref{eqn:ShiftedfreqNthOrderDyson}. 

\section{Multiple Drive Inputs}\label{sec:ArbDriveInputs}

Suppose there are~$q$-independent drive inputs each with own drive frequency and operator. For an~$n$-th order Dyson expansion, we define an $n$-dimensional vector~$\bm{\beta}$ with values ranging from~$1$ to~$q$, each value referring to one of the drive inputs. To simplify the  notation, we drop all subscripts $n$, which are implied. The new Dyson series operators corresponding to a particular vector $\bm{\beta}$ is
\begin{equation}
\begin{aligned}
 &\hat{S}_{\bm{\beta}}^{(n)}(\bm{\omega}_{\bm{\beta}},\delta t) = \sum_{\bm{k}_n} \langle k^{(n)} |\hat{X}_{\bm{\beta}[n]}|k^{(n-1)}\rangle \langle k^{(n-1)} |...\langle k^{(1)} |\hat{X}_{\bm{\beta}[1]} | k^{(0)} \rangle | k^{(n)} \rangle \langle k^{(0)} | f(\bm{\lambda} -\delta t\bm{v}_{\bm{\beta}} ),
\end{aligned}
\end{equation}
with $\bm{\omega}_{\bm{\beta}}$ an $n-$vector where the $p$-th element of this vector is $\pm \omega_{\bm{\beta}[p]}$, or more simply, plus or minus the drive frequency corresponding to the $\bm{\beta}[p]$-th input. Similarly, $\hat{X}_{\bm{\beta}[p]}$ refers to the $\bm{\beta}[p]$-th input operator. The definition of $v_{\bm{\beta}}$ remains unchanged from \cref{eqn:Freqvector}, albeit with the new drive frequency vector $\bm{\omega}_{\bm{\beta}}$. The drive function is also modified and now reads
\begin{equation} \label{eqn:DriveFunction}
\begin{aligned}
	\Omega(\bm{\omega}_{\bm{\beta}}) &= \prod_{p=1}^n \bm{\Omega}_{\bm{\beta}[p]}^{\mu} \bm{\Omega}_{\bm{\beta}[p]}^{* (1-\mu)},
\end{aligned}
\end{equation}
where 
\begin{equation}
\begin{aligned}
	\mu = \frac{1}{2}\big( 1+\textrm{sign} \lbrace\bm{\omega}_{\bm{\beta}}[p] \rbrace \big),
\end{aligned}
\end{equation}
thus allowing us to define our generic~$n$-th order Dyson series,
\begin{equation}
\begin{aligned}
\hat{U}^{(n)}(t,t+\delta t) = \sum_{\bm{\beta}} \sum_{\bm{\omega}_{\bm{\beta}}}  \exp(i\sum_{p=1}^n  \bm{\omega_\bm{\beta}} [p]t)\Omega(\bm{\omega}_{\bm{\beta}}) \hat{S}_{\bm{\beta}}^{(n)}(\bm{\omega}_{\bm{\beta}},\delta t),
\end{aligned}
\end{equation}
which reads as a version of \cref{eqn:FullUnitaryNthOrdersingledrive} with multiple drive operators and frequencies.

\section{Linear interpolation of subpixels}
\label{app:LinearInterpolation}

\begin{figure}
	\centering
	\includegraphics[width=0.5\columnwidth]{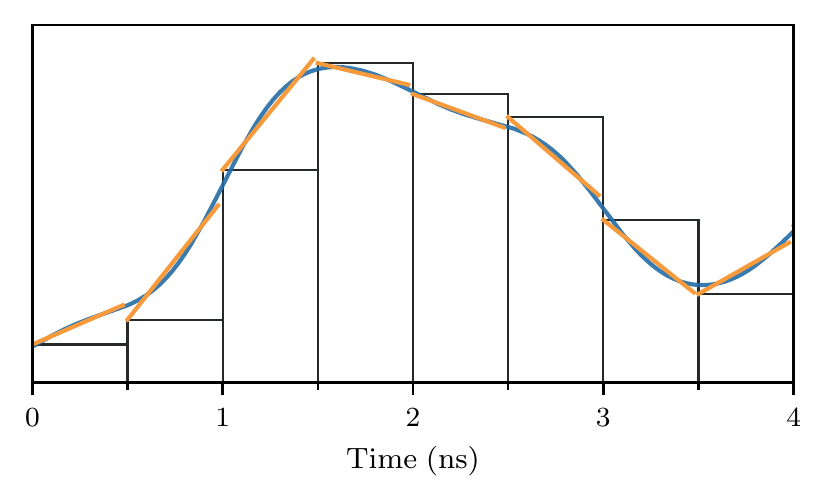}
	\caption{Linear interpolation of the pulse sequences (yellow) with two subpixels per pixel. The black bars indicate the original Gaussian filtering, with the blue line the `true' filtered pulse.}
	\label{fig: INterpolated-pulse}
\end{figure}

The accuracy of the Dyson expansion is insensitive to the drive frequency $\omega$ -- rather, it depends only on the number of subpixels and the drive amplitudes. However, if each subpixel assumes a constant amplitude, a leading error can be caused by the change in amplitude of the drive over a single subpixel. In \cref{fig: INterpolated-pulse}, we illustrate this issue, where the subpixel amplitudes (black bars) will over or underestimate the true pulse amplitude, depending on the gradient of the envelope function. To circumvent this issue, we consider a new linear interpolation of the drive envelope over a single subpixel. We define $y_l(t)$ as the new time dependent amplitude for the $l$-th subpixel:
\begin{equation}\label{eqn:linearinterpdrive}
\begin{aligned}
	y_l(t) &= s'_l + \frac{s_{l+1} - s_l}{\delta t}(t-l\delta t), \quad l\delta t < t < (l+1)\delta t, \\
\end{aligned}
\end{equation}
where $s'_l$ is calculated from a modified filter function to ensure that the integral of the subpixel matches that of the continuous pulse.  The $y_l(t)$ time-dependent subpixels are shown in orange, and well approximate the true pulse even for relatively few subpixels.

To determine the impact of the linear time term $t_0$ on the time ordered integral, we consider the $(n-j)$-th iteration of a Dyson Series operator,

\begin{equation}
	\begin{split}
		&\int_0^{\delta t}dt_M\cdots\int_0^{t_{M-m_{(j+1)}}}\hspace{-3em}dt_{M-m_j}  t_{M-m_{j}} \exp(i\bm{\omega}_n[j]t_{M-m_{j}})\hat{S}^{(j-1)}(\bm{g}^{(j-1)}(\bm{\omega}_{n}),t_{M-m_{j}})\,,\\
&=\int_0^{\delta t}dt_M\cdots\int_0^{t_{M-m_{(j+1)}}}\hspace{-3em}dt_{M-m_j} \left( -i \partial_{\bm{\omega}_n,j} \right) \exp(i\bm{\omega}_n[j]t_{M-m_{j}})\hat{S}^{(j-1)}(\bm{g}^{(j-1)}(\bm{\omega}_{n}),t_{M-m_{j}})\,,\\
&= \left( -i \partial_{\bm{\omega}_n,j} \right)\int_0^{\delta t}dt_M\cdots\int_0^{t_{M-m_{(j+1)}}}\hspace{-3em}dt_{M-m_j}  \exp(i\bm{\omega}_n[j]t_{M-m_{j}})\hat{S}^{(j-1)}(\bm{g}^{(j-1)}(\bm{\omega}_{n}),t_{M-m_{j}})\,,\\
&= \left( -i \partial_{\bm{\omega}_n,j} \right) \hat{S}^{(j)}(\bm{g}^{(j)}(\bm{\omega}_{n}),t_{M-m_{j+1}})\,,\\
	\end{split}
\end{equation}
where we define $\partial_{\bm{\omega}_n,j}$ as the derivative with respect to the $j$-th component of the vector $\bm{\omega}_n$. For example
\begin{align}
    \partial_{\bm{\omega}_n,j}(\bm{\omega_n}) = [0,0,\cdots,\underbrace{1}_\text{j-th},\cdots,0].
\end{align}
This result allows us to modify \cref{eqn:DriveFunctioneasy} to the following result, replacing the drive amplitude function by a drive amplitude operator,
\begin{equation}
\hat{\Omega}_l(\bm{\omega}_n) = \prod_{p=1}^n \left( s_l -i s_l'\partial_{\bm{\omega}_n,p} \right)^{\mu} \left( s_l^* +is_l'^* \partial_{\bm{\omega}_n,p}\right)^{1-\mu}, \textrm{ where }
\mu = \frac{1}{2}\big( 1+\textrm{sign} \lbrace\bm{\omega}[p] \rbrace \big),
\end{equation}
where the derivatives are to act upon the Dyson series operators. It is important to note that $s_l' \ll s_l$ in the majority of cases -- as such, it is possible to consider a partial truncation of the series, where only up to a certain power of derivative terms $s_l'$ are included in the set of Dyson series operators. 

To make use of the drive amplitude function, we must determine the derivatives of the weighting function $f(\bm{\lambda}_n)$. We begin with the base case with $\bm{\lambda}_0 = \left[\lambda_0\right]$,
\begin{equation}
	\begin{split}
		\frac{d}{d\lambda_0}f(\bm{\lambda}_0) &= \frac{d}{d\lambda_0}e^{-i\lambda_0}\\
		&= -i e^{-i\lambda_0}\\
		&= \lim_{\epsilon \rightarrow 0} \frac{e^{-i(\lambda_0 + \epsilon)} - e^{-i\lambda_0}}{\epsilon}\\ 
		&= if(\lambda_0 \cup \bm{\lambda}_0 ).
	\end{split}
\end{equation}
Now, we assume that $\partial_{\bm{\lambda}_n,j} f(\bm{\lambda}_n) = if(\bm{\lambda}_{n}[j] \cup \bm{\lambda}_n)$. Then,
\begin{equation}
\begin{split}
    \partial_{\bm{\lambda}_n,j}f(\bm{\lambda}_{n}) &= i\partial_{\bm{\lambda}_n,j} \frac{f(g(\bm{\lambda}_{n})) - f(g^2(\bm{\lambda}_{n}) \cup \bm{\lambda}_{n}[n])}{\bm{\lambda}_{n}[n-1] - \bm{\lambda}_{n}[n]}\\
    &= \frac{i}{\bm{\lambda}_{n}[n-1] - \bm{\lambda}_{n}[n]} \left[\partial_{\bm{\lambda}_n,j}f(g(\bm{\lambda}_{n})) - \partial_{\bm{\lambda}_n,j}f(g^2(\bm{\lambda}_{n}) \cup \bm{\lambda}_{n}[n]) \right]\\
        &= \frac{i}{\bm{\lambda}_{n}[n-1] - \bm{\lambda}_{n}[n]} \left[ if(g(\bm{\lambda}_{n}[j] \cup \bm{\lambda}_{n})) - if(g^2(\bm{\lambda}_{n}[j] \cup \bm{\lambda}_{n}) \cup \bm{\lambda}_{n}[n]) \right]\\
        &= i f(\bm{\lambda}_{n}[j] \cup \bm{\lambda}_{n}).
    \end{split}
\end{equation}

\section{Benchmark calculations \label{app:Benchmark}}

\begin{figure}
	\centering
	\includegraphics[width=1.\columnwidth]{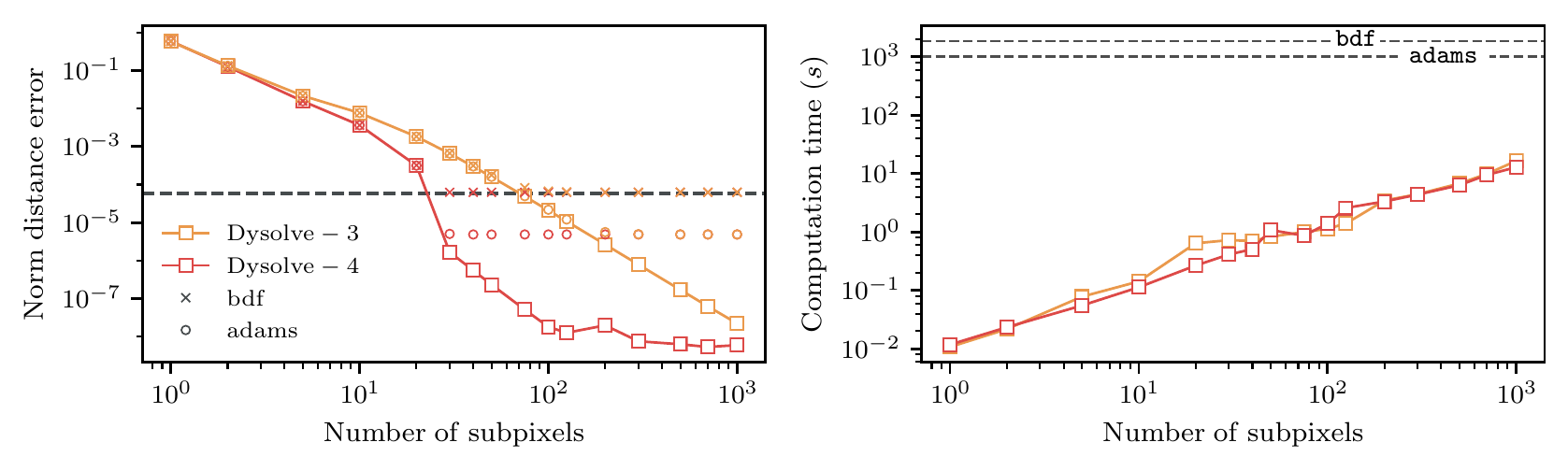}
	\caption{Comparison of the \texttt{Dysolve} algorithm and QuTiP's \texttt{propagator} over an evolution of $20$ Rabi oscillations, with the same parameters as in Fig \ref{fig: benchmark_6panels} of the main text.
	\textbf{Left panel}: Frobenius norm distance between the \texttt{Dysolve} algorithm with different subpixel numbers and QuTiP's \texttt{propagator} with  \texttt{adams} (circles) and \texttt{bdf} (x's) methods. The squares are the norm distance between \texttt{Dysolve-n} and \texttt{Dysolve-4} with $10^4$ subpixels. The dashed line indicates the error between \texttt{propagator} with the \texttt{adams} and \texttt{bdf} methods.
	\textbf{Right panel}: Computational time of \texttt{Dysolve} and \texttt{propagator} for results of the left panel. The dashed lines refer to the computational time for \texttt{propagator} with \texttt{adams} and \texttt{bdf} with the highly accurate settings described in the text.}
	\label{fig: DysolveBenchmarkvsPropagator2panels}
\end{figure}

In order to calculate the Frobenius norm error used as a metric in our benchmarks, we required an excellent approximation to the reference propagator operator $\hat{U}_{\mathrm{ref}}(0,T)$. We first considered an alternative numerical solver, \texttt{propagator} from the \texttt{python} package QuTiP.
We selected highly accurate settings for this computation, namely absolute and relative tolerances of $10^{-16}$, 2 different numerical methods with their maximal order ($12$ for \texttt{adams} and $5$ for \texttt{bdf}), discretizing each pixel into $10^4$ subpixels and allowing for $10^{17}$ internal sub-steps.
After averaging over many simulations of $T=500$~ns with a single drive, it became apparent that the \texttt{propagator} function was converging and unable to return solutions with a Frobenius norm error less than $5\times10^{-6}$, as seen in \cref{fig: DysolveBenchmarkvsPropagator2panels}.
The shortcoming of this numerical method was confirmed by the fact that the norm distance between \texttt{adams} and \texttt{bdf} was found to be significantly greater than the norm distance between \texttt{adams} and the \texttt{Dysolve} algorithm for a sufficient subpixel number.
This is in addition to a computational time at least 2 order of magnitudes larger with the standard QuTiP approach as seen in the right panel of \cref{fig: DysolveBenchmarkvsPropagator2panels}.

\end{document}